\documentclass[aps,prl,nobibnotes,twocolumn,superscriptaddress,bibliography]{revtex4-2}
\usepackage{amsfonts}
\usepackage{mathrsfs}
\usepackage{amsmath}
\usepackage{color}
\usepackage{natbib}
\usepackage{graphicx}
\usepackage{bm}
\usepackage{amssymb}
\usepackage{xspace}
\usepackage{epstopdf}
\usepackage{dcolumn}
\usepackage{longtable}
\usepackage{array}
\usepackage{makecell}
\usepackage{tabulary}
\usepackage{multirow}
\usepackage{appendix}
\usepackage{mathtools}
\usepackage{pifont}
\usepackage{longtable}
\newcolumntype{C}[1]{>{\centering\arraybackslash}p{#1}}
\newcolumntype{L}[1]{>{\flushleft\arraybackslash}p{#1}}

\usepackage{multirow}
\usepackage{epsfig}
\usepackage[normalem]{ulem}
\usepackage{amsmath}
\usepackage{braket}
\usepackage{bbold}
\usepackage{graphicx}
\usepackage{xcolor}

\usepackage[normalem]{ulem}

\makeatletter

\usepackage[colorlinks=true, letterpaper=true, pdfstartview=FitV, linkcolor=blue, citecolor=blue, urlcolor=blue]{hyperref}

\makeatother
\begin{document}

\title{Two-Dimensional Spin-Antiferroelectric Altermagnets with Giant Spin Splitting: From Model to Material Realization}

\author{Zesen Fu}
\thanks{These authors contributed equally to this work.}
\affiliation{School of Physics and Technology, Xinjiang University, Urumqi 830017, China}
\affiliation{School of Physics, Central South University, Changsha 410083, China}

\author{Aolin Li}
\thanks{These authors contributed equally to this work.}
\affiliation{School of Physics and Technology, Xinjiang University, Urumqi 830017, China}

\author{Wenzhe Zhou}
\affiliation{School of Physics, Central South University, Changsha 410083, China}

\author{Fangping Ouyang}
\email{ouyangfp06@tsinghua.org.cn}
\affiliation{School of Physics and Technology, Xinjiang University, Urumqi 830017, China}
\affiliation{School of Physics, Central South University, Changsha 410083, China}

\author{Fawei Zheng}
\email{fwzheng@bit.edu.cn}
\affiliation{Beijing Key Lab of Nanophotonics and Ultrafine Optoelectronic Systems, School of Physics, Beijing Institute of Technology, Beijing, China.}

\author{Yugui Yao}
\email{ygyao@bit.edu.cn}
\affiliation{Beijing Key Lab of Nanophotonics and Ultrafine Optoelectronic Systems, School of Physics, Beijing Institute of Technology, Beijing, China.}

\date{\today}

\begin{abstract}
The realization of multiferroic altermagnets featuring giant intrinsic spin splitting, hold great promise for next-generation spintronics.
In this work, based on the recently proposed concept of spin-antiferroelectric (spin-AFE), we construct a class of two-dimensional (2D) multiferroic altermagnets, termed 2D spin-antiferroelectric altermagnets (2D spin-AFEAMs), enabling electrical control of spin polarization via a gate field.
Furthermore, we propose a general design strategy for constructing 2D spin-AFEAMs with large intrinsic spin splitting. 
Guided by this strategy, we predict monolayer $(\mathrm{CoCl})_2\mathrm{Te}$ and its family materials as potential candidates of 2D spin-AFEAM. 
We uncover a highly tunable transport regime in monolayer $(\mathrm{CoCl})_2\mathrm{Te}$, where the spin current can be switched via the in-plane electric field angle when hole-doped, and via the gate field polarity when electron-doped. 
Our work enriches the family of 2D multiferroics and provides a blueprint for realizing high-performance, electrically switchable altermagnetic spintronic devices.
	
\end{abstract}

\maketitle

\textit{Introduction.}---Multiferroics, characterized by the coexistence of magnetic and (anti)ferroelectric ((A)FE) orders, possess intrinsic magnetoelectric coupling~\cite{MF0,MF1,MF2}.
and provide a versatile platform for manipulating electronic properties via external magnetic and electric fields~\cite{MF0,MF1,MF2,MF3,MF4,MF5,MF6}.
However, FE states require insulating behavior, which is often incompatible with the metallic nature of ferromagnetism~\cite{MF7}.
Consequently, realizing multiferroicity by combining antiferromagnet (AFM) with (A)FE state has emerged as a feasible and widely adopted strategy~\cite{MF8,MF9,MF10,MF11,MF12}.
Nevertheless, the application of such AFM-based multiferroics is severely restricted by a fundamental limitation: conventional AFMs typically exhibit spin-degenerate bands (lack of spin polarization), hindering their utility in spintronic devices~\cite{AFM2,AFM1}.

The emergence of altermagnets (AMs) offers a promising solution to this dilemma~\cite{AM2,AM1,AM3,AM4}. 
AMs are an emergent class of magnetic materials that combine properties of ferromagnets and antiferromagnets, exhibiting spin-polarized band structures and zero net magnetic moment due to combined time-reversal and crystal symmetry~\cite{AM2,AM1,AM3,AM4,2014japC,2019jpsjS,2020prbL,2021prlR,AM5,AM6,AM7,AM8,AM9,AM10}. 
Current research on AMs has explored a wide array of exotic phenomena, including spin-polarized currents~\cite{AM4,sc1,sc2,sc3,sc4,sc5,sc6}, nontrivial superconductivity~\cite{super1,super2,super3}, piezomagnetism~\cite{pm1}, anomalous Hall effect~\cite{ahc1,ahc2,ahc3,ahc4}, and other intriguing spintronic behaviors~\cite{AlterWeyl,2025prlC,2025prlL,2025aF,2025np1,2025np2,2024prlY}.

Recently, the intersection of altermagnetism and multiferroicity has attracted significant theoretical attention. 
Investigations have focused on various coupling mechanisms, including the coupling between AM order and (anti)ferroelectricity~\cite{FEAM1,FEAM2,FEAM3,AFEAM1}, fractional quantum ferroelectricity~\cite{FQFE1,FQFE2}, the altermagnetoelectric effect~\cite{AME}, and Type-II antiferroelectricity~\cite{II-AFE}. 
However, a recurrent limitation in these reported multiferroic AM candidates is their relatively weak spin splitting. 
This challenge motivates a fundamental question: Is there a general strategy to significantly enhance the difference of spin, or to rationally design multiferroic AMs that inherently possess large intrinsic spin splitting?

In this paper, we address these challenges by building on the recently proposed concept of spin-antiferroelectric altermagnets (spin-AFEAMs~\cite{II-AFE}), which are defined by the decoupling of distinct spin subspaces with equal magnitude and opposite orientation polarizations.
By replacing the ill-defined Berry connection under open boundary conditions with the position operator, we extend spin-AFEAMs to 2D systems and establish the corresponding symmetry requirements.
We further demonstrate that 2D spin-AFEAMs inherently exhibit magnetoelectric coupling , and the spin polarization can be efficiently controlled by a gate field via spin-resolved band structure modulation.
Furthermore, we formulate a rational design strategy to engineer 2D spin-AFEAMs with large intrinsic spin splitting. 
Guided by this strategy, we predict the class of candidate materials: monolayer $(A_2\mathrm{Cl}_2)X$ ($A = \mathrm{Cr, Mn, Fe, Co, Ni}$; $X = \mathrm{O, S, Se, Te}$). 
Notably, monolayer $(\mathrm{CoCl})_2\mathrm{Te}$ exhibits a unique dual-control regime for spin transport: spin currents can be manipulated by the in-plane electric field direction under hole doping, whereas they are controlled by the gate field under electron doping. 
Our work not only enriches the family of 2D multiferroics but also provides a comprehensive roadmap for designing high-performance altermagnetic spintronic devices.

\textit{Definition of 2D Spin-AFEAM}---
Building on the definition of spin-AFE~\cite{II-AFE}, we consider an insulating altermagnet with a collinear magnetic configuration. In the absence of spin-orbit coupling (SOC), the electronic Hilbert space $V$ can be decoupled into orthogonal spin-up and spin-down sectors, 
$V = V_{\uparrow} \oplus V_{\downarrow}$.
Consequently, the Bloch Hamiltonian block-diagonalizes as:
\begin{equation}
	\label{eq:1}
	\mathcal{H}(\bm{k})=H_{\uparrow}(\bm{k})\oplus H_{\downarrow}(\bm{k}).
\end{equation}
To calculate the polarization for each spin sector within the Berry-phase framework~\cite{spin-AFE1,spin-AFE2,spin-AFE3} and explicitly treating the 2D limit (open boundary conditions along the $z$ direction), we use the expectation value of the position operator $\hat{z}$ rather than the Berry connection. And
\begin{equation}
	\label{eq:3}
	\hat{z} = \sum_n z_n |n\rangle \langle n|,
\end{equation}
where $|n\rangle$ represents the $n$-th tight-binding basis orbital located at $z_n$ and the summation includes all orbitals.
Thus, the $z$-direction polarization in the 2D limit is given by:
\begin{equation}
	\label{eq:4}
	P^{z}_{\uparrow/\downarrow} = -\frac{e}{4\pi^2} \sum_{n}^{\mathrm{occ}} \int_{\mathrm{BZ}} \langle n_{\uparrow/\downarrow}(\bm{k}) | \hat{z} | n_{\uparrow/\downarrow}(\bm{k}) \rangle \, dk_x dk_y.
\end{equation}
Here, the summation is over all occupied states, and $|n_{\uparrow/\downarrow}(\bm{k})\rangle$ denotes the eigenfunction of the $n$-th band at momentum $\bm{k}$. 
A system is defined as a 2D spin-AFEAM if these spin-resolved polarizations are equal in magnitude but opposite in sign, i.e.,
\begin{equation}
	\label{cdt-AFEAM}
	P^{z}_{\uparrow}=-P^z_{\downarrow}\neq0.
\end{equation}

Here, we demonstrate the symmetry constraints.
For an AM with negligible SOC, the elements of its nontrivial spin group can be universally categorized into the following two forms\cite{AM1,AM2,AM3}:
\begin{equation}
	\label{eq:5}
\left[C_2\right| \left|R\right], \quad \left[E\right|\left| R^\prime\right].
\end{equation}
In this notation, the operations on the left of
the double vertical bar acts only in spin space,($C_2$ being the spin-flip operation and $E$ the identity), while those on the right of the double
vertical bar simultaneously act on the real space. 
Here, $R$ and $R^\prime$ represent spatial point group elements, where $R$ specifically connects opposite magnetic sublattices.

The operation $\left[C_2\right| \left|R\right]$ interchanges the spin indices while acts upon the spatial operator $\hat{z}$. 
Consequently, based on Eq.~(\ref{eq:4}), to preserve this symmetry alongside the spin-AFEAM condition in Eq~\eqref{cdt-AFEAM}, the spatial transformation $R$ is required to reverse the $z$-direction (i.e., $z \to -z$). 
Conversely, the operation $R^\prime$ must preserve the sign of $z$. 
Finally, to satisfy the fundamental criteria of altermagnetism, $R$ must not correspond to spatial inversion $\mathcal{P}$ or translation $\tau$.

\textit{Properties of 2D Spin-AFEAM}---
Spin-AFEAMs inherently exhibit magnetoelectric coupling,
and tunable spin polarization via gate field.

Considering a four-band model comprising one valence band and one conduction band for each spin channel. In the absence of SOC, the tight-binding Hamiltonian can be expressed in the following form:
\begin{equation}
	\label{eq:6}
	\mathcal{H}=d_0+d_x\tau_x+d_y\tau_y+d_z\tau_z+u\tau_z\sigma_z.
\end{equation}
Here, $\bm{\tau}$ and $\bm{\sigma}$ denote the Pauli matrices acting on the sublattice and spin degrees of freedom, respectively.
The coefficients $d_{i}$ (with $i=0, x, y, z$) are real functions of momentum $\bm{k}$.
The term $u\tau_z\sigma_z$ describes the magnetic exchange coupling on the distinct sublattices. 
Notably, this term is odd under sublattice exchange or time reversal; thus, $u$ serves as the order parameter for the antiferromagnetic moment.

To compute the spin-resolved polarization $P^z_{\uparrow/\downarrow}$, we analyze the position operator $\hat{z}$ from Eq.~\eqref{eq:3}. 
Decomposed into the Pauli basis, it takes the form:
\begin{equation}
	\label{eq:7}
	\hat{z} = z_0 + z_1\tau_z + z_2\sigma_z + z_3\tau_z\sigma_z.
\end{equation}
Applying the symmetry constraints derived previously, we note that $R$ (It is from Eq.~\eqref{eq:5}) dictates $\hat{z} \to -\hat{z}$, $\tau_z \to -\tau_z$ and $d_z(\bm{k})=-d_z(R^{-1}\bm{k})$. 
Additionally, $\hat{z}$ must be even under time-reversal symmetry. 
These conditions collectively enforce $\hat{z} = z_1\tau_z$. 
Using Eq.~(\ref{eq:4}), we then arrive at:
\begin{equation}
	\label{eq:8}
	\begin{split}
	p^{z}_{\uparrow/\downarrow}(\bm{k})&=\sum_{n \in \mathrm{occ}}\langle n_{\uparrow/\downarrow}(\bm{k})|\hat{z}|n_{\uparrow/\downarrow}(\bm{k})\rangle=\frac{d_z\pm u}{\sqrt{d_x^2+d_y^2+d_z^2}}, \\
	P^{z}_{\uparrow/\downarrow}&=-\frac{ez_1}{4\pi^2}\int_{\mathrm{BZ}}p^{z}_{\uparrow/\downarrow}(\bm{k})dk_xdk_y=\pm\alpha u,
	\end{split}
\end{equation}
here, $\alpha=-\frac{ez_1}{4\pi^2}\int_{\mathrm{BZ}}\frac{1}{|\bm{d}|}dk_xdk_y$.
This result reveals that the spin-AFE order parameter is directly proportional to the antiferromagnetic order parameter. 
Such a linear dependence explicitly demonstrates the magnetoelectric coupling within the system.

\begin{figure}
	\centering 
	\includegraphics[width=0.5\textwidth]{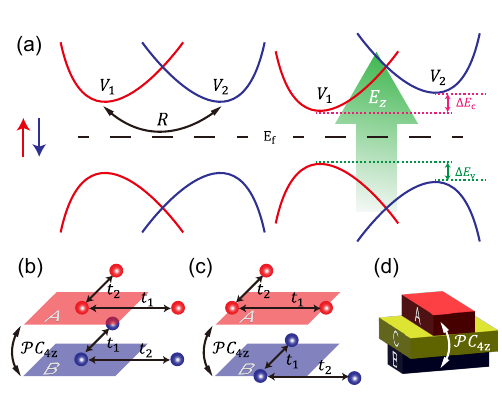}%
	\caption{\label{fig:1} 
		(a) Schematic illustration of the spin-AFEAM band structure in the absence and presence of a gate field. 
		In the unperturbed state, two degenerate valleys $V_1$ and $V_2$ in the conduction (valence) band with opposite spin are related by the symmetry operation $R$, reflecting the intrinsic altermagnetic symmetry. 
		Upon applying a gate field, energy shifts $\Delta E_c$ and $\Delta E_v$ are induced with opposite signs, leading to contrasting variations in the band gaps for opposite spin channels.
		Design strategy for constructing 2D spin-AFEAMs demonstrated in (b), (c) and (d).
		The structure consists of a top Layer A (red), a bottom Layer B (blue), and an intermediate Layer C (yellow). 
		Red and blue spheres represent magnetic atoms with opposite moments. 
		Layers A and B are related by the $\mathcal{P}C_{4z}$ symmetry operation, with $t_1$ and $t_2$ denoting distinct anisotropic hopping integrals ($t_1 \neq t_2$). 
		Panels (b) and (c) illustrate two different stacking configurations: (b) with vertically aligned magnetic atoms, and (c) with spatially staggered magnetic atoms. 
		Panel (d) demonstrates the strategy for giant intrinsic spin splitting, where the intermediate Layer C supports the highly anisotropic Layers A and B.}
\end{figure}

\begin{figure}
	\centering 
	\includegraphics[width=0.5\textwidth]{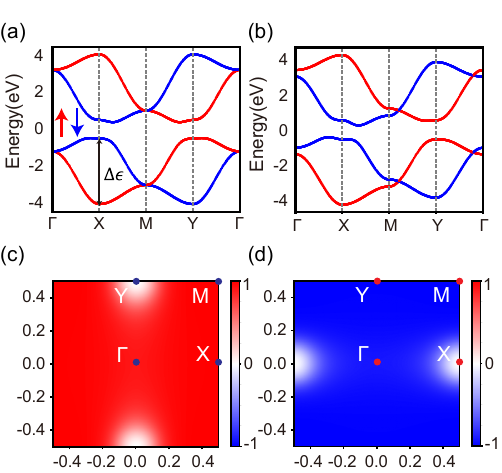}%
	\caption{\label{fig:2} 
		Calculated band structures for the AA-stacked configuration in the (a) absence and (b) presence of a gate field. 
		(c) and (d): Momentum-space distribution of the spin-resolved polarization density, $p^z_{\uparrow/\downarrow}(\bm{k}) = \sum_{n \in \mathrm{occ}}\langle n_{\uparrow/\downarrow}(\bm{k})|\hat{z}|n_{\uparrow/\downarrow}(\bm{k})\rangle$, for the (c) spin-up and (d) spin-down channels. Note that the calculated polarization values have been normalized for visual clarity.
		The model parameters are set to $t_1=-0.25$, $t_2=0.75$, $\mu=0$, $u=2$, and $t_\bot=-0.5$.(eV) 
		Based on Eqs.~\eqref{eq:7} and \eqref{eq:9}, the gate field perturbation is modeled as $-eE_z z_1\tau_z$, with the field strength parameter fixed at $-eE_z z_1 = 0.2$ eV.}
\end{figure}

Here, we analyze the response of the 2D spin-AFEAM to a gate field. 
Within the perturbation theory framework, the perturbation Hamiltonian takes the form:

\begin{equation}
	\label{eq:9}
	H^{\prime} = -e E_z \hat{z},
\end{equation}
where $E_z$ denotes the electric field strength along the $z$-direction. 
Focusing on the spin-up sector, the first-order energy correction is given by $\epsilon^{(1)}_{n,\uparrow}(\bm{k}) = -eE_z \langle n_{\uparrow}(\bm{k}) | \hat{z} | n_{\uparrow}(\bm{k}) \rangle$. 
Considering the summation over all bands, we find:
\begin{equation}
	\begin{split}
		\sum_n \epsilon^{(1)}_{n,\uparrow}(\bm{k}) &= \sum_{n \in \mathrm{occ}} \epsilon^{(1)}_{n,\uparrow}(\bm{k}) + \sum_{n \in \mathrm{unocc}} \epsilon^{(1)}_{n,\uparrow}(\bm{k}) \\
		&= -eE_z \mathrm{Tr} \left( \sum_n |n_{\uparrow}(\bm{k})\rangle \langle n_{\uparrow}(\bm{k})| \hat{z} \right) = 0.
	\end{split}
\end{equation}
Here, the subscripts ``occ" and ``unocc" denote summation over occupied and unoccupied states, respectively. 
The vanishing trace implies that an upward shift in the spin-up valence bands is necessarily accompanied by a downward shift in the corresponding conduction bands.

\begin{figure*}
	\centering 
	\includegraphics[width=1.0\textwidth]{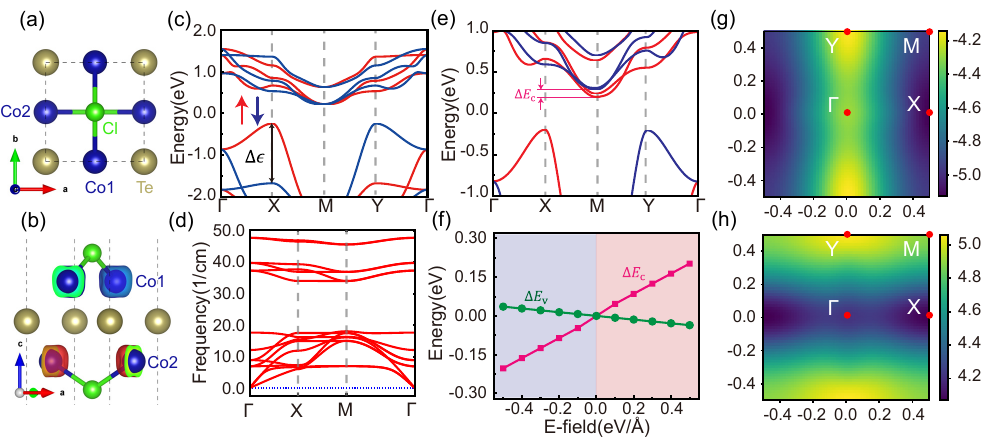}%
	\caption{\label{fig:3} 
		(a) and (b): Atomic structure of monolayer $(\mathrm{CoCl})_2\mathrm{Te}$. 
		Blue, green, and yellow atoms represent Co, Cl, and Te, respectively. 
		Panel (b) visualizes the spin density difference $\Delta\rho = \rho_{\uparrow} - \rho_{\downarrow}$. 
		Red and blue regions correspond to positive ($\Delta\rho > 0$) and negative ($\Delta\rho < 0$) values, respectively, with color intensity indicating the magnitude. 
		(c) and (e): Electronic band structures calculated noSOC in the (c) absence and (e) presence of a gate field ($E_z=0.2\,\mathrm{eV/\AA}$). 
		Red and blue lines denote spin-up and spin-down channels, respectively. 
		(d) Calculated phonon spectrum of monolayer $(\mathrm{CoCl})_2\mathrm{Te}$. 
		(f) Spin splitting for CBM ($\Delta E_{c}$, pink line) and VBM ($\Delta E_{v}$, green line) as a function of the gate field. 
		The red (blue) shaded background indicates the regime where the spin-up (spin-down) band gap is smaller.
		(g) and (h): Momentum-space distribution of the spin-resolved polarization density, $p^z_{\uparrow/\downarrow}(\bm{k}) = \sum_{n \in \mathrm{occ}}\langle n_{\uparrow/\downarrow}(\bm{k})|\hat{z}|n_{\uparrow/\downarrow}(\bm{k})\rangle$, for the (g) spin-up and (h) spin-down channels.
	}
\end{figure*}

Furthermore, the operations $\left[C_2\right| \left|R\right]$ interchange the spin indices while transforming $\hat{z} \to -\hat{z}$. 
Thus, the relation will get:
\begin{equation}
	\label{eq:11}
	\sum_{n \in \mathrm{occ}} \epsilon^{(1)}_{n,\uparrow}(R\bm{k}) = -\sum_{n \in \mathrm{occ}} \epsilon^{(1)}_{n,\downarrow}(\bm{k})=\sum_{n \in \mathrm{unocc}} \epsilon^{(1)}_{n,\downarrow}(\bm{k}).
\end{equation}
This indicates that the gate field induces opposite energy shifts for the spin-up and spin-down bands.
		
This mechanism is illustrated in Fig.~\ref{fig:1}(a). In the unperturbed spin-AFEAM state, the conduction band features two degenerate valleys, $V_1$ and $V_2$, with opposite spin polarizations linked by a crystalline symmetry $R$. 
An applied gate field breaks this symmetry and induces a downward energy shift in $V_1$. 
Crucially, the intrinsic constraints of spin-AFEAM (Eq.~\eqref{eq:11}) compel $V_2$ to shift upward, while valence bands of the same spin experience opposite energy shifts. 
Consequently, the spin-up band gap narrows while the spin-down gap widens. 
This dictates that the spin splittings at the conduction band minimum (CBM), $\Delta E_c$ and the valence band maximum (VBM), $\Delta E_v$ exhibit opposite signs.
Moreover, reversing the polarity of the electric field swaps these effects between the spin channels. 
Therefore, spin-AFEAM systems provide a robust platform for realizing electrically tunable spin polarization.

\textit{General Strategy for Realizing 2D Spin-AFEAM}---
Here, we propose a strategy for designing 2D spin-AFEAMs exhibiting giant spin splitting. We construct a minimal model to illustrate the physical properties, confirming the theoretical predictions derived above.

Consider a tetragonal lattice system illustrated in Fig.~\ref{fig:1}(b). 
Layer A (red) is a ferromagnetic monolayer lacking $C_{4z}$ symmetry. 
Layer B (blue) is generated from Layer A via $\mathcal{P}C_{4z}$, an operation combining spatial inversion with a $90^\circ$ rotation. 
Imposing antiferromagnetic coupling between the layers and containing $[C_2||\mathcal{P}C_{4z}]$ symmetry, thereby realizing a 2D spin-AFEAM.

Our minimal model features a unit cell with one magnetic site per layer. For Layer A, we define anisotropic nearest-neighbor (NN) hopping amplitudes, $t_1$ and $t_2$, along the $x$ and $y$ axes, respectively. 
Layer B is rotated by $90^\circ$ relative to Layer A. 
Thus, the corresponding tight-binding Hamiltonian (only consider NN) is given by:

\begin{equation}
	\label{eq:12}
	\begin{split}
		\mathcal{H} &= \mu + (t_1+t_2)(\cos{k_x}+\cos{k_y}) \\
		& + (t_1-t_2)(\cos{k_x}-\cos{k_y})\tau_z + f(t_{\perp})\tau_x + u\tau_z\sigma_z.
	\end{split}
\end{equation}
Here, $\mu$ is the on-site energy, $t_{\perp}$ the interlayer coupling, and $u$ the exchange coupling associated with the AFM order. The interlayer hopping $f(t_{\perp})$ depends on the stacking configuration. 
For simple AA stacking (Fig.~\ref{fig:1}(b)), $f(t_{\perp})=t_{\perp}$ is a constant. 
However, for the configuration in Fig.~\ref{fig:1}(c), $f(t_{\perp})= 4t_{\perp}\cos(k_x/2)\cos(k_y/2)$. 

Fig.~\ref{fig:2}(a) shows the calculated band structure, featuring characteristic altermagnetic splitting.
Figs.~\ref{fig:2}(c) and (d) map the spin-resolved polarization density, $p^z_{\uparrow/\downarrow}(\bm{k}) = \sum_{n \in \mathrm{occ}}\langle n_{\uparrow/\downarrow}(\bm{k})|\hat{z}|n_{\uparrow/\downarrow}(\bm{k})\rangle$, across the Brillouin zone. 
The distributions for the two spin channels are opposite in sign, strictly satisfying the $[C_2||\mathcal{P}C_{4z}]$ symmetry and confirming the realization of a 2D spin-AFEAM.
Furthermore, applying an external gate field (Fig.~\ref{fig:2}(b)) narrows the spin-up band gap while widening the spin-down gap.
This directly validates our theoretical model, demonstrating that the spin-AFEAM system enables gate-tunable spin polarization.

A recurrent challenge in multiferroic altermagnets is their relatively weak spin splitting. Our minimal model provides a clear design principle to overcome this limitation.
At the X point [Fig.~\ref{fig:2}(a)], the energy difference between the highest occupied states of opposite spins characterizes the spin splitting magnitude:
\begin{equation}
	\Delta\epsilon =\frac{4uB}{\sqrt{(2B+u)^2+f^2(t_\perp)}+\sqrt{(2B-u)^2+f^2(t_\perp)}}.
\end{equation}
Here, $B=t_1-t_2$ represents the in-plane anisotropy. Crucially, this gap widens as $f(t_\perp)\to0$, demonstrating that the stacking configuration in Fig.~\ref{fig:1}(c) yields a larger spin splitting than simple AA stacking.
In this stacking, $\Delta\epsilon \propto (t_1 - t_2)$, dictating that Layer A must exhibit strong in-plane anisotropy to maximize the intrinsic splitting.

Fig.~\ref{fig:1}(d) illustrates the new design strategy. We select a strongly anisotropic Layer A, such as a 1D atomic chain, and generate Layer B via the $\mathcal{P}C_{4z}$ operation. Because the direct stacking of Layers A and B is generally unstable, an intermediate Layer C is inserted to stabilize the lattice. This resulting sandwich architecture realizes a 2D spin-AFEAM featuring giant spin splitting.

\textit{Material realization}---
Guided by the design strategy above, we demonstrate a family of monolayer $(A_2\mathrm{Cl}_2)X$ materials ($A = \mathrm{Co, Cr, Fe, Mn, Ni}$; $X = \mathrm{O, S, Se, Te}$)  are the potential candidates of 2D spin-AFEAM with giant spin splitting. In the following, we focus on monolayer $(\mathrm{CoCl})_2\mathrm{Te}$ as a representative material. The other candidate materials are provided in the Supplemental Material~\cite{SM}.

\begin{figure}
	\centering 
	\includegraphics[width=0.5\textwidth]{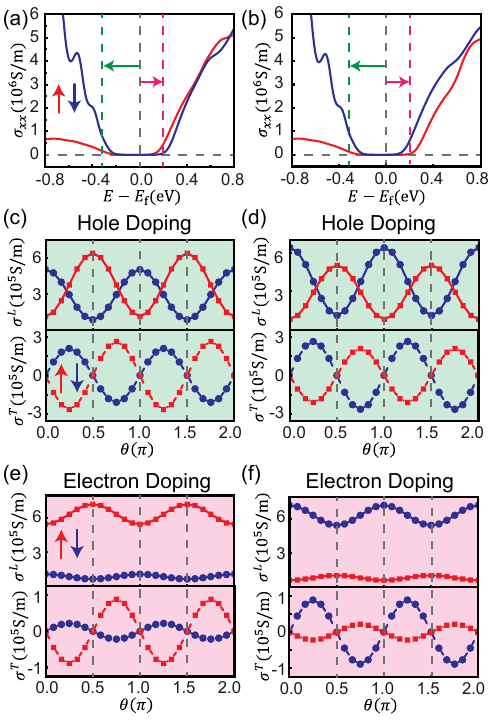}%
	\caption{\label{fig:4} 
(a) Calculated Boltzmann conductivity as a function of energy at room temperature ($300\,\mathrm{K}$) under a gate field of $E_z=0.2\,\mathrm{eV/\AA}$. 
Red and blue curves denote the spin-up and spin-down channels, respectively. 
The green and pink arrow indicate hole doping ($0.012~\mathrm{e/u.c.}$) and electron doping ($0.078~\mathrm{e/u.c.}$), respectively. 
(c) Angular dependence of the longitudinal ($\sigma^L$) and transverse ($\sigma^T$) conductivities for the hole-doped case, where $\theta$ represents the angle between the in-plane electric field and the $+x$ axis. 
(e) Corresponding conductivity plot for the electron-doped case. 
(b), (d) and (f): Equivalent results calculated under a reversed gate field of $E_z=-0.2\,\mathrm{eV/\AA}$.
	}
\end{figure}

Figs.~\ref{fig:3}(a) and (b) illustrate the atomic structure of monolayer $(\mathrm{CoCl})_2\mathrm{Te}$. 
The spin-down Co atoms and their surrounding Cl ligands form 1D chains with pronounced in-plane anisotropy, acting as the effective top layer (Layer A). 
The bottom layer (Layer B), containing spin-up Co atoms, is generated from Layer A via the $\mathcal{P}C_{4z}$ operation. 
Finally, Te atoms located at the unit cell corners serve as the intermediate layer (Layer C) to stabilize the sandwich structure.

As shown in Fig.~\ref{fig:3}(c), the material also exhibits significant spin splitting ($\Delta\epsilon=1.44$eV).
To further verify the spin-AFEAM nature, we map the momentum-space distribution of the spin-resolved polarization, $p^z_{\uparrow/\downarrow}(\bm{k}) = \sum_{n \in \mathrm{occ}}\langle n_{\uparrow/\downarrow}(\bm{k})|\hat{z}|n_{\uparrow/\downarrow}(\bm{k})\rangle$, in Figs.~\ref{fig:3}(g) and (h). 
The distributions display opposite signs for the two spin channels and strictly obey the $[C_2||\mathcal{P}C_{4z}]$ symmetry, confirming the antiparallel alignment of the spin-dependent electric polarization. 

We further investigated the electronic structure under a field of $E_z=0.2\,\mathrm{eV/\AA}$ (Fig.~\ref{fig:3}(e)). 
The valence bands show a negligible response, whereas the conduction bands exhibit a significant shift. 
Fig.~\ref{fig:3}(f) displays the evolution of $\Delta E_{c/v}$ under varying gate fields. 
The observed linear dependence and the opposite signs of the energy shifts are in excellent agreement with our theoretical model.

Exploiting the distinct sensitivities of the valence and conduction bands to the electric field, we propose a strategy for the active control of spin currents. 
Figs.~\ref{fig:4}(a) and (b) present the calculated Boltzmann conductivity $\sigma_{xx}$ at room temperature ($300\,\mathrm{K}$) under gate fields of $E_z = 0.2$ and $-0.2\,\mathrm{eV/\AA}$. 
We assume that the Fermi level can be tuned to specific regimes via light carrier doping. Thus, The green and pink dashed line denote the corresponding Fermi level under hole and electron doping, respectively.

We investigate the transport properties by plotting the longitudinal ($\sigma^L$) and transverse ($\sigma^T$) conductivities as a function of the in-plane electric field angle $\theta$ (defined relative to the $+x$ axis). 
Taking the hole doping at $E_z = 0.2\,\mathrm{eV/\AA}$ (Fig.~\ref{fig:4}(c)) as a representative example, we observe a reversal in the longitudinal spin current: the dominance shifts from $\sigma^L_{\downarrow} > \sigma^L_{\uparrow}$ to $\sigma^L_{\uparrow} > \sigma^L_{\downarrow}$ as the field direction rotates from $\theta=0$ to $\pi/2$. 
Crucially, due to the insensitivity of the valence band to the vertical field, the dominant spin channel remains unchanged even when the gate field polarity is reversed (Fig.~\ref{fig:4}(d)).

The scenario is fundamentally different for electron doping. 
Fig.~\ref{fig:4}(e) displays the conductivities for the electron-doped case under $E_z = 0.2\,\mathrm{eV/\AA}$. 
Here, we observe $\sigma^L_{\uparrow} > \sigma^L_{\downarrow}$. Reversing the gate field polarity inverts this relationship to $\sigma^L_{\uparrow} < \sigma^L_{\downarrow}$. 
Based on these findings, we conclude that the system offers a versatile control mechanism: in the hole-doped regime, the spin current is manipulated by the in-plane electric field direction (and is robust against the gate field), whereas in the electron-doped regime, it is controlled by the gate field (and is largely insensitive to the in-plane field direction).

\textit{Conclusion}---
In summary, we have established the theoretical foundation for 2D spin-AFEAMs, demonstrating that these systems host intrinsic magnetoelectric coupling and gate-tunable spin polarization properties. 
Furthermore, we proposed a general design strategy for constructing 2D spin-AFEAMs, identifying the key criteria to maximize intrinsic spin splitting. 
Applying this strategy, we identified the $(A_2\mathrm{Cl}_2)X$ family as robust 2D spin-AFEAM candidates with giant spin splitting. 
Notably, in monolayer $(\mathrm{CoCl})_2\mathrm{Te}$, we uncovered a dual-control mechanism for spin transport: in the hole-doped regime, spin currents are manipulated by the in-plane electric field orientation and are robust against the vertical gate field; conversely, when electron-doped, spin currents are controlled by the gate field and remain insensitive to the in-plane field direction. 
Our work establishes spin-AFEAMs as a promising platform for developing advanced spintronic devices with versatile control functionalities.

\textit{Discussion}---
While this work has primarily focused on the spin transport properties of spin-AFEAMs, a plethora of other physical properties of spin-AFEAMs remain to be uncovered in future research.
Furthermore, although we demonstrated our design protocol using tetragonal lattices via the $\mathcal{P}C_{4z}$ symmetry, the underlying principle is not limited to this specific geometry. 
The framework can be naturally adapted to other crystal systems, such as triangular lattices, where $\mathcal{P}C_{6z}$ symmetry can be invoked to realize analogous spin-AFEAM structures.
The present definition of spin-AFEAM relies on the assumptions of collinear magnetism, weak SOC, and an insulating ground state. 
Generalizing this concept to overcome these constraints awaits further investigation.

\textit{Acknowledgments}---
This work was financially supported by the Key Project of the Natural Science Program of Xinjiang Uygur Autonomous Region (Grant No. 2023D01D03), the National Natural Science Foundation of China (Grant No. 52073308, No. 12304097 and No. 12164046), the Tianchi-Talent Project for Young Doctors of Xinjiang Uygur Autonomous Region (No. 51052300570) and the State Key Laboratory of Powder Metallurgy at Central South University. This work was carried out in part using computing resources at the High Performance Computing Center of Central South University.

\bibliography{letter1.bib}

\end{document}